# Obstacle-Free Path Planning for Autonomous Drones Using Floyd's Algorithm

Edward Yao
*Walton High School, Marietta, GA, USA*
*edwardyao@gatech.edu*

**Abstract**
*This research investigates the efficiency of Floyd's algorithm for obstacle-free path planning for autonomous aerial vehicles (UAVs) or drones. Floyd's algorithm is used to generate the shortest paths for UAVs to fly from any place to the destination in a large-scale field with obstacles which UAVs cannot fly over. The simulation results demonstrated that Floyd's algorithm effectively plans the shortest obstacle-free paths for UAVs to fly to a destination. It is verified that Floyd's algorithm holds a time complexity of $O(n^3)$. This research revealed a correlation of a cubic polynomial relationship between the time cost and the size of the field, no correlation between the time cost and the number of obstacles, and no correlation between the time cost and the number of UAVs in the tested field. The applications of the research results are discussed in the paper as well.*

## I. INTRODUCTION

Unmanned aerial vehicles (UAVs), also known as drones, are aircraft that operate without a human pilot or passengers onboard. More often UAVs are controlled remotely by a human pilot although they can be fully or partially autonomous. Recent research and development on these drones have led to drones being used in many areas [1], including agricultural [2] and logistical applications [3], monitoring and surveillance [1]. Drones have been used for both military and commercial [4] to monitor and locate targets in complex terrain. They play important roles in rescue operations. UAVs can supply goods to people in inaccessible locations. In the military, drones can be used to survey the land and locate targets to be attacked. Especially, with the rise of threats to humanity and conflicts, having the ability to help rescue people and find landmarks without the risk of loss of human life is crucial.

Because of the increasing number and complexity of tasks that UAVs are expected to perform, it is crucial to plan an optimal path for UAVs to fly to a location of interest. The shortest obstacle-free path will help UAVs to fly to the designated site to perform their tasks quickly and efficiently. Researchers have applied existing algorithms to find the shortest and optimal path for UAVs to perform tasks in human inaccessible areas. The most researched algorithm is the Dijkstra algorithm, which finds the shortest path from one source to a destination [5, 6, 7]. However, in numerous cases, multiple UAVs housed in different places of the same region are engaged in the same service, making it imperative to pick the UAV which takes the least time to fly from its current location to the target site to perform the urgent tasks. This needs a computation and comparison of the shortest paths of all the UAVs from their resting site to the target point. Floyd's algorithm, an algorithm to find the shortest paths from each vertex in a weighted graph to each of the other vertices, is an excellent fit to solve the problems where multiple UAVs are available for a task however only the one which can arrive at the target point most quickly should be picked.

In this research, we will study the application of Floyd's algorithm to path planning for UAVs. The path generated from this research is the shortest path that avoids obstacles to UAVs reaching a specific point or site. The effectiveness of applying Floyd's algorithm in obstacle-free path planning will be evaluated.

## II. PATH PLANNING AND FLOYD'S ALGORITHM

Path planning is a process of finding an optimal sequence of points which move an object from a starting point to the end point along the path. In this research, the path planning will find the shortest path for UAVs to use the least time to fly from one point to the target point in a field. In addition, if there are multiple UAVs in service and only one is needed, the UAV which arrives at the target point fastest will be identified to perform the task. The input factors are starting point and target point of an UAV if only one UAV is considered, starting points of UAVs and the target point of UAVs if multiple UAVs are involved, and number and locations of obstacles. The output will be a shortest 2-D path from the start point to the target point for a UAV which avoids obstacles if one UAV is involved. If multiple UAVs are involved the output will be the set of the shortest path from each UAV's start point to the target point. The UAV which arrives at the target point most quickly is identified to perform a task at the target point.

Floyd's algorithm, shortened from Floyd-Warshall algorithm, is an algorithm that is best suited to find the shortest paths between all pairs of nodes in weighted graphs. Unlike Dijkstra algorithm, Floyd's algorithm works for both the directed and undirected weighted graphs and handles both positive and negative edge weights of the graphs. Floyd's algorithm works by creating a matrix with entries w(i, j) representing the edge weights (distance) between all pairs of nodes $i$ and $j$ of a graph. The matrix is initialized with the direct edge weights between the nodes. If there is no direct edge between nodes $i$ and $j$, the value is set to infinity, ∞. Then, for each vertex $k$, the algorithm checks if the path from vertex $i$ to vertex $j$ through vertex $k$ is shorter than the current path from vertex $i$ to vertex $j$. If so, the matrix updates that entry of the matrix. This is repeated for all vertices. The result will be the matrix with the shortest path distances between all pairs of vertices.

To plan the shortest path for a UAV to fly from a starting point to a destination point in a field which is divided into grids of the same size, an initial weight matrix is created based on the visibility and direct distance between any pair of the grids. Two grids are not visible to each other if there is one or more obstacles lying between them, this means a drone cannot fly directly from one grid to the other one because the obstacle(s) blocks the way for the drone to make the fly. The drone must use other grids and take turns to fly to a grid which is not visible to the starting grid. The weight or distance between two invisible grids is infinity meaning it is not possible to reach one grid from another one by a straight line flying. If two grids are visible to each other, then the distance between them is the straight distance meaning a UAV can fly from one of the grids directly to another one along a straight-line path. Figure 1 illustrates a field of 16 grids with 4 obstacles which are in grids 2, 7, 8, and 10, respectively. As shown in Figure 1, grid 1 and grid 5 are visible to each other, the distance between them is 1, grid 1 and grid 9 are visible to each other, the distance between them is 2, similarly, the distance between grid 1 and grid 13 is 3. However, grid 1 and grid 3 are not visible because there is an obstacle on grid 2 between them, the distance between them is ∞, similarly the distance between grid 1 and grid 4 is ∞. Grid 6 and grid 11 are not visible to each other because grid 7 and grid 10 are corner touched which makes it impossible for a UAV to fly from grid 6 to grid 11 and vice versa. The distance between grid 6 and grid 11 is ∞. The distance between grid 5 and grid 14 is ∞ because an obstacle exists on grid 10 which blocks a drone flyting from grid 5 to grid 14. The distance between grid 12 and grid 14 is √5. The weighted, directed graph with 16 vertices generated from Figure 1 is shown in Figure 2. The weight on each edge in Figure 2 is the direct straight distance between two grids if they are visible to each other. The weighted matrix $W$ (Figure 3) representing the weighed graph in Figure 2 is generated where

$$W[i][j] = \begin{cases} \text{weight on edge} & \text{if there is an edge from } vi \text{ to } vj \\ \infty & \text{if there is no edge from } vi \text{ to } vj \\ 0 & \text{if } i = j \end{cases}$$

Pseudo code for Floyd's Algorithm for path planning is shown in Figure 4.

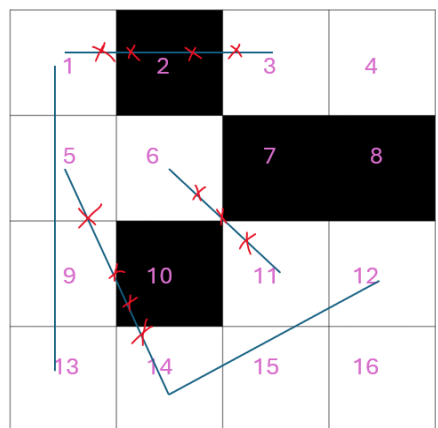

**Figure 1:** A field with 16 unit-grids demonstrating the visibility of pairs of grids

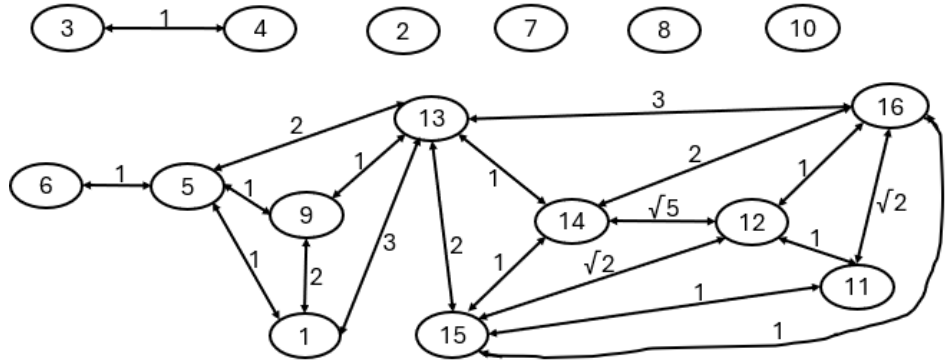

**Figure 2:** The weighted directed graph generated from the field in Figure 1.

|    | 1 | 2 | 3 | 4 | 5 | 6 | 7 | 8 | 9 | 10 | 11 | 12 | 13 | 14 | 15 | 16 |
|----|---|---|---|---|---|---|---|---|---|----|----|----|----|----|----|----|
| 1  | 0 | ∞ | ∞ | ∞ | 1 | ∞ | ∞ | ∞ | 2 | ∞  | ∞  | ∞  | 3  | ∞  | ∞  | ∞  |
| 2  | ∞ | 0 | ∞ | ∞ | ∞ | ∞ | ∞ | ∞ | ∞ | ∞  | ∞  | ∞  | ∞  | ∞  | ∞  | ∞  |
| 3  | ∞ | ∞ | 0 | 1 | ∞ | ∞ | ∞ | ∞ | ∞ | ∞  | ∞  | ∞  | ∞  | ∞  | ∞  | ∞  |
| 4  | ∞ | ∞ | 1 | 0 | ∞ | ∞ | ∞ | ∞ | ∞ | ∞  | ∞  | ∞  | ∞  | ∞  | ∞  | ∞  |
| 5  | 1 | ∞ | ∞ | ∞ | 0 | 1 | ∞ | ∞ | 1 | ∞  | ∞  | ∞  | 2  | ∞  | ∞  | ∞  |
| 6  | ∞ | ∞ | ∞ | ∞ | 1 | 0 | ∞ | ∞ | ∞ | ∞  | ∞  | ∞  | ∞  | ∞  | ∞  | ∞  |
| 7  | ∞ | ∞ | ∞ | ∞ | ∞ | ∞ | 0 | ∞ | ∞ | ∞  | ∞  | ∞  | ∞  | ∞  | ∞  | ∞  |
| 8  | ∞ | ∞ | ∞ | ∞ | ∞ | ∞ | ∞ | 0 | ∞ | ∞  | ∞  | ∞  | ∞  | ∞  | ∞  | ∞  |
| 9  | 2 | ∞ | ∞ | ∞ | 1 | ∞ | ∞ | ∞ | 0 | ∞  | ∞  | ∞  | 1  | ∞  | ∞  | ∞  |
| 10 | ∞ | ∞ | ∞ | ∞ | ∞ | ∞ | ∞ | ∞ | ∞ | 0  | ∞  | ∞  | ∞  | ∞  | ∞  | ∞  |
| 11 | ∞ | ∞ | ∞ | ∞ | ∞ | ∞ | ∞ | ∞ | ∞ | ∞  | 0  | 1  | ∞  | ∞  | 1  | √2 |
| 12 | ∞ | ∞ | ∞ | ∞ | ∞ | ∞ | ∞ | ∞ | ∞ | ∞  | 1  | 0  | ∞  | √5 | √2 | 1  |
| 13 | 3 | ∞ | ∞ | ∞ | 2 | ∞ | ∞ | ∞ | 1 | ∞  | ∞  | ∞  | 0  | 1  | 2  | 3  |
| 14 | ∞ | ∞ | ∞ | ∞ | ∞ | ∞ | ∞ | ∞ | ∞ | ∞  | √5 | 1  | 0  | 1  | 2  |
| 15 | ∞ | ∞ | ∞ | ∞ | ∞ | ∞ | ∞ | ∞ | ∞ | ∞  | 1  | √2 | 2  | 1  | 0  | 1  |
| 16 | ∞ | ∞ | ∞ | ∞ | ∞ | ∞ | ∞ | ∞ | ∞ | ∞  | √2 | 1  | 3  | 2  | 1  | 0  |

**Figure 3:** Matrix *W* representing the graph in Figure 2.

Input: Weighted matrix *W[n][n]*, where *n* is the number of grids in a field
Output: Shortest distance matrix *D[n][n]*, where *n* is the number of grids in a field. This
matrix saves the shortest distance between any two grids in the field.
Shortest path matrix *P[n][n]* where *n* is the number of grids in the field. This
matrix records the vertices on the shortest path of a pair of vertices.
Floyd (input *W*, output *D, P*)
Initiate *P = 0*
*D = W*
for *k* in *n*
   for *i* in *n*
      for *j* in *n*
         if $D_{ik} + D_{kj} < D_{ij}$ then
            $P_{ij} = k$
            $D_{ij} = D_{ik} + D_{kj}$

**Figure 4:** Pseudo code of Floyd's algorithm.

### III. SIMULATION SETTINGS

Several scenarios are designed to study the effectiveness of Floyd' algorithm in planning optimal obstacle-free paths for UAVs in a field with an arbitrary number of obstacles. First the simulation will demonstrate that the shortest path can be generated by using Floyd's algorithm for a drone which flies from a source to the destination in a field with an arbitrary number of obstacles. Then we will test how size of field or number of grids in the field, number of obstacles in the field, number of starting points in the field are correlated to the performance of Floyd's algorithm in terms of time cost.

Scenario 1: The number of obstacles, number of sources, and the size of the field are constant. This setting is to demonstrate that Floyd's algorithm generates the shortest path for UAVs to fly from one place to another in a field.
- Size of field: 100 (10×10) or 3,600 (60×60) grids
- Number of obstacles: 20 on the 10×10 field and 100 on the 60×60 field
- Number of source vertices: 1 or multiple

Scenario 2: The number of obstacles and number of sources are constant while the size of the field is changing. This is to test how size of field (number of grids) affects the performance of path planning using Floyd's algorithm.
- Size of field: 10×10 (100), 20×20 (400), 30×30 (900), 40×40 (1600), 50×50 (2500), 60×60 (3600), 70×70 (4900), 80×80 (6100), 90×90 (8100), and 100×100 (10000).
- Number of obstacles: 100
- Number of source vertices: 1
- Performance measurements: time cost

Scenario 3: The size of the field and number of sources are constant while the number of obstacles is changing. This is to test how the number of obstacles affects the performance of path planning using Floyd's algorithm in a field.
- Size of field: 3600 (60×60)
- Number of obstacles: 1, 100, 200, 300, 400, 500, 600, 700, 800, 900, and 1000.
- Number of source vertices: 1
- Performance measurements: time cost

Scenario 4: The size of the field and number of obstacles are constant while the number of sources is changing. This is to test the correlation between the number of UAVs which can serve on the same task in a field and the time cost.
- Size of field: 3600 (60×60)
- Number of obstacles: 100
- Number of source vertices: 1, 5, 10, 20, 50,100, 200, 400, 500, 600, and 1000.
- Performance measurements: time cost

## IV. SIMULATION RESULTS AND DISCUSSION

Figure 5 demonstrates the shortest paths generated by applying Floyd's algorithm on a field of 10×10 grids with 20 obstacles and a field of 60×60 with 100 obstacles. Figure 5a demonstrates a case where there is no route available for a UAV to fly from the source grid to the destination grid because there are obstacles blocking the way on a 2-D field. Figure 5 (b), (c), and (e) show that a shortest path from the source to the destination is calculated and planned by Floyd's algorithm on two fields of different size and number of obstacles. Figure 5 (d) and (f) demonstrate the shortest paths from multiple source grids to a single destination. As shown in Figure 5, it is obvious that Floyd's algorithm can be used to find the shorted path from a starting point to a target point. However, if there is only one UAV available and it is housed in a place from which there is no path to the destination, the task or incident in the destination will not have an opportunity to be handled (Figure 5a). Floyd's algorithm calculates the shortest paths between any pairs of vertices in a graph, which can be used to manage the cases where multiple UAVs are available for the service but are housed in different locations (Figure 5 d, f). In these cases, Floyd's algorithm generates the shortest path for each UAV to the target point, the UAV which is closest to the destination should be tasked to fly to the target destination. This management ensures that the target point will receive the service of a UAV in the shortest time.

The time cost is highly correlated to the size of field holding a polynomial relationship as shown in Figure 6. The regression is $y = 6*10^{-10}x^3 + 3*10^{-6} x^2 - 0.0074$, $R^2 = 0.995$. This result confirms the time complexity of $O(n^3)$ for Floyd's algorithm. The size of field needs to be considered when applying Floyd's algorithm in path planning. Due to the cubic relationship between the time cost and the size of field, an appropriate size of fields should be recommended depending on extent of urgency of the task in the target grid. If the task is not urgent, then the size of the field could be larger because the time cost for generating the shortest path does not affect the total performance of the task. However, if the incident in the target grid is urgent, the shortest path needs to be calculated quickly, a smaller size of field should be recommended.

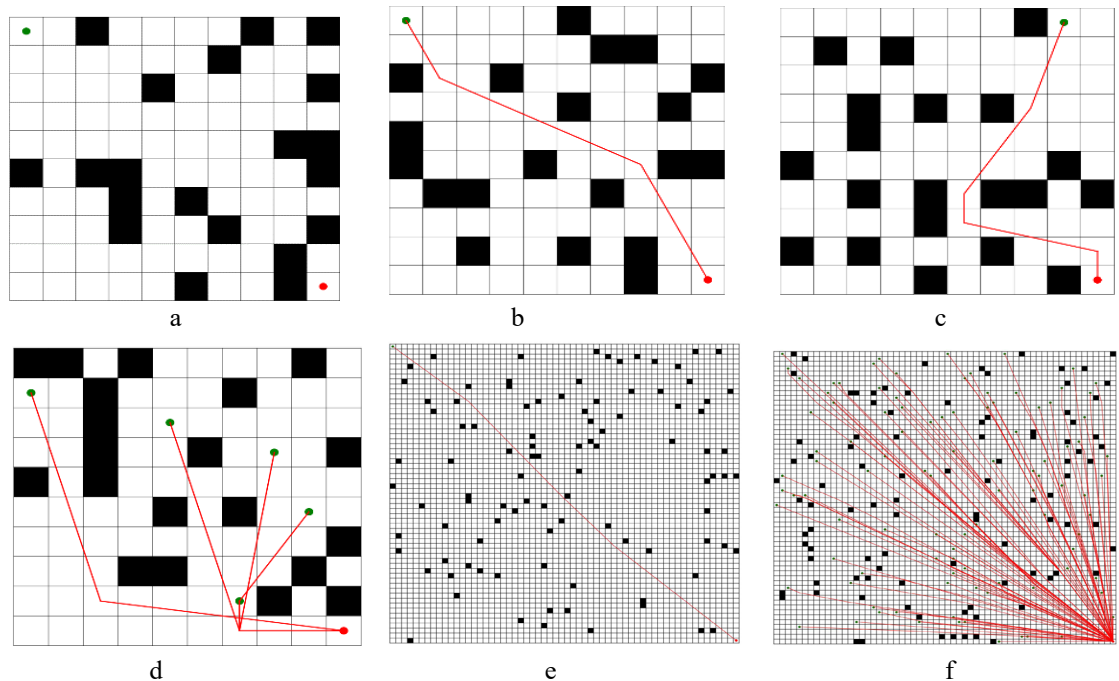

**Figure 5:** Shortest path from source (green dots) to destination (red dot) planned by Floyd's algorithm on a 10×10 grids field with 20 obstacles or 60×60 grids field with 100 obstacles. (a) shows that there is no route from the source to the destination; (b)-(f) show the shortest path from the source(s) to the destination.

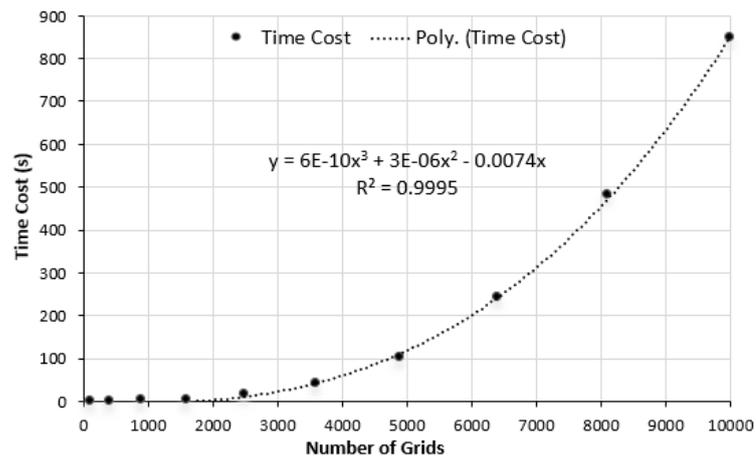

**Figure 6:** The cubic relationship between time cost and the size of fields.

The experiment shows that there is no relationship between time cost and the number of obstacles in the fields (Figure 7). The flat line and low $R^2$ value (0.0532) in Figure 7 indicate that no matter how many obstacles exist in a field, the time cost for Floyd's algorithm to generate the shortest paths for all the pairs of grids in the field is constant.

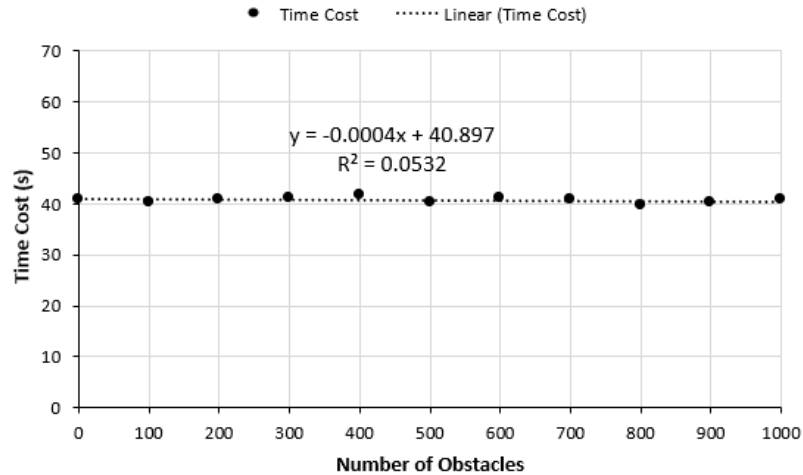

**Figure 7:** Relationship between time cost and number of obstacles in a field of 60×60 grids

It is shown in Figure 8 that there is no correlation between the time cost for Floyd's algorithm to generate the shortest paths from an arbitrary number of starting points to the target point and the number of starting points. According to the simulation results, the $R^2$ for a linear regression between time cost and the number of sources is significantly low, $R^2 = 7*10^{-5}$. The Floyd's algorithm, unlike Dijkstra algorithm, calculates the shortest distance between any pair of vertices on a graph. No matter how many UAVs which are housed in different places in a field are available for a task, Floyd's algorithm always generates the shortest path from any UAV in the field to the destination. For urgent response, it is suggested to prepare multiple UAVs in the service area. Once there is an incident in a specific place, Floyd's algorithm can calculate and plan the shortest path for each UAV from its current location to the destination. By comparing the shortest paths generated for these UAVs, the UAV which is closest to the destination can be easily identified and ordered to fly to the destination and perform tasks.

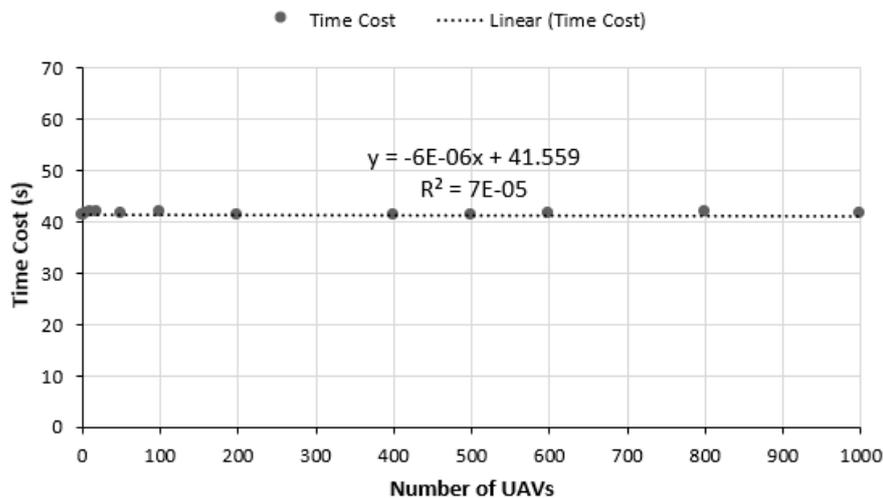

**Figure 8:** Relationship between time cost and number of UAVs in a field of 60×60 grids

## V. CONCLUSION

This research investigated the efficiency of Floyd's algorithm for optimal obstacle-free path planning for autonomous UAVs. The specific purpose of this research is to study the effects of the size of the field, the number of obstacles, and the number of UAVs on the performance of Floyd's algorithm in planning shortest obstacle-free paths for UAVs. The results demonstrated that Floyd's algorithm effectively plans optimal obstacle-free paths for UAVs to fly to a destination, especially when there are multiple UAVs at different locations. The algorithm can help find the UAV which is closest to the destination and should take the task to fly to the destination. This research verified that the time complexity of Floyd's algorithm is $O(n^3)$, where n is the number of vertices in the graph. The simulation results revealed a very high correlation between the time cost and the size of the field with a cubic polynomial relationship ($6*10^{-10}x^3 + 3*10^{-3}x^2 - 0.0074x$, $R^2 = 0.9995$ ), no correlation between the time cost and the number of obstacles in the field ($R^2 = 0.0532$), and no correlation between the time cost and the number of source UAVs ($R^2 = 7*10^{-5}$).